\documentclass[manuscript]{aastex}
\begin{document}
\title{Synthetic Spectral Analysis of the Hot Component
   in the S-Type Symbiotic Variable EG Andromeda}

\author{Kelly Kolb}
\affil{Dept. of Astronomy \& Astrophysics,
Villanova University,
Villanova, PA 19085,
email: kelly.kolb@villanova.edu}

\author{Joleen Miller}
\affil{Dept. of Astronomy \& Astrophysics,
Villanova University,
Villanova, PA 19085,
email: joleen.miller@villanova.edu}

\author{Edward M. Sion}
\affil{Dept. of Astronomy \& Astrophysics,
Villanova University,
Villanova, PA 19085,
e-mail: edward.sion@villanova.edu}

\author{Joanna Mikolajewska}
\affil{Copernicus Astronomical Center,
Polish Academy of Sciences, Warsaw, Poland,
e-mail: mikolaj@camk.edu.pl}

\begin{abstract} 
We have applied grids of NLTE high gravity model
atmospheres and optically thick accretion disk models for the first time
to archival IUE and FUSE spectra of the S-type symbiotic variable EG And
taken at superior spectroscopic conjunction when Rayleigh scattering
should be minimal and the hot component is viewed in front of the red
giant. For EG And's widely accepted, published hot component mass, orbital
inclination and distance from the Hipparcos parallax, we find that hot,
high gravity, NLTE photosphere model fits to the IUE spectra yield
distances from the best-fitting models which agree with the Hipparcos
parallax distance but at temperatures substantially lower than the
modified Zanstra temperatures.  NLTE fits to an archival FUSE spectrum
taken at the same orbital phase as the IUE spectra yield the same
temperature as the IUE temperature (50,000K). However, for the same hot
component mass, inclination and parallax-derived distance, accretion disk
models at moderately high inclinations, $\sim 60-75^{\circ}$ with
accretion rates $\dot{M} = 1\times 10^{-8}$ to $1\times 10^{-9}
M_{\sun}/yr$ for white dwarf masses $M_{wd} = 0.4 M_{\sun}$ yield
distances grossly smaller than the distance from the Hipparcos parallax.
Therefore, we rule out an accretion disk as the dominant source of the FUV
flux. Our findings support a hot bare white dwarf as the dominant source
of FUV flux. 
\end{abstract}

Subject Headings: Stars: symbiotic variables, white dwarfs, Physical
Processes: accretion, accretion disks, wind accretion

\section{Introduction}

Symbiotic stars are binary systems typically consisting of a red giant
primary with a hot white dwarf companion orbiting close enough to the red
giant that it can accrete material from the giant's wind. Most symbiotic
systems contain a normal red giant and these, based on their near-IR
colours (consistent with the presence of a stellar atmosphere with
T$_{eff}\ 3000 - 4000$K) are classified as S-type systems (stellar). The
remainder ($\leqq 20$\,\%) contain Mira variables and their near-IR
colours are consistent with the combination of a reddened Mira and a warm
dust shell: these are classified as D-type systems (dusty).

In many symbiotic systems, there is little or no evidence of an accretion
disk surrounding the hot, accreting component. In fact, there is no strong
evidence of a disk in any S-type symbiotic containing a white dwarf. This
also appears to be true for D-type symbiotics. There is also no evidence
for a FUV contribution by a nebular continuum shortward of about 1600 \AA.
Therefore the FUV continuum may be contributed virtually entirely by the
hot accreting white dwarf radiation up to 1800 \AA. The presence of a hot
accreting white dwarf in symbiotic binaries offers the possibility of
studying the effect of wind accretion on the compact star and to further
test whether accretion disks surround the accreting compact component.
This is especially important in view of the smooth particle hydrodynamic
simulations of Mastrodemos and Morris (1998) which predict disk formation
from wind accretion. In their 3D wind accretion models, they found that
permanent accretion disks form in detached binaries with a variety of
separations including those characteristic of the binary properties of
symbiotic variables. Their disks lead to significant accretion of mass and
angular momentum by the mass-gaining component and some of the the disk
properties conform to those expected from steady state disk theory
(Mastrodemos and Morris 1998).

The synthetic spectra of hot, high gravity stellar atmospheres and
optically thick steady-state accretion disks offer the possibility of
directly determining the properties of the hot component and further
testing for the presence or absence of an accretion disk. {\it Up to now,
there has been no direct comparison of the FUV spectra of the hot
components of symbiotics with realistic models of hot white dwarf
atmospheres and accretion disk models with vertical structure}. We have
selected EG And as a followup to our study of the hot component in RW
Hydrae (Sion et al. 2002).  What are the model-derived properties of the
accreting hot component in EG And and how do its properties compare with
the results of black body fits and Zanstra techniques?

\section{Observed Parameters of the EG And System}

We have selected EG Andromedae as a followup study to RW Hydrae (Sion et
al. 2002). EG And exhibits a sharp rise in the FUV continuum of IUE
spectra indicating the presence of a hot component.  The observed
parameters for EG And from the literature are compiled in Table 1.  EG And
has a Hipparcos parallax which yields a distance of 676 pc (see below).

Vogel et al. (1992) estimated the red giant radius, $R_{\rm g} = 74 \pm
10\, \rm R_{\odot}$, and the system inclination,
$i=82^{\circ}\stackrel{+8}{-4.5}$, from analysis of the UV eclipse light
curve and the Rayleigh scattering effects. They also found the stellar
masses, $M_{\rm g} = 1.5\, \rm M_{\odot}$ and $M_{\rm h} = 0.4\, \rm
M_{\odot}$, respectively. The low mass of the red giant is consistent with
the observed $\rm C^{12}/C^{13} \sim 10$ (Schild et al. 1992), typical for
red giants with masses of $ 2-3 \rm M_{\odot}$.  The upper limit for
$M_{\rm g} <3\, \rm M_{\odot}$ combined with the mass function derived
from the cool giant absorptions (e.g. Belczy{\'n}ski et al. 2000) implies
the upper limit for $M_{\rm h} <0.6\, \rm M_{\odot}$.

The estimate for the giant radius of Vogel et al. is distance independent.
Wilson \& Vaccaro (1997) analyzed ellipsoidal variability in the optical
range, and found that the giant fills $82-85\,\%$ of its Roche lobe. Using
the binary parameters from Vogel et al. (1992) this implies the red giant
radius is $128 - 133\, \rm R_{\odot}$, and the distance, $d=670$ pc in
excellent agreement with the Hipparcos parallax.

\begin{deluxetable}{lcr}
\tablewidth{0pc} 
\tablecaption{EG And System Parameters}
\tablehead{
\colhead{Parameter}&

\colhead{Value}&
\colhead{Reference}  }
\startdata
Orbital Period (d)      &      482.6            & 1                     \\
JD$_{min}$              &      2445380            & 1                           \\
E(B-V)                  &            0.05         & 1                   \\            
Hipparcos parallax (mas)&      $1.48\pm0.97$     & 2                            \\
Distance (pc)           &      $676^{+1285}_{-268}$    &\nodata \tablenotemark{a}\\
Inclination             &      $78-90^{\circ}$    & 3                   \\
Spec Type Cool          &      M3 III           & 1                     \\
M$_{cool}$  ($M\sun$)   &      $ 1.5$           & 3                     \\
R$_{cool}$  ($R\sun$)   &      $75 \pm 10$     & 3                      \\
T$_{eff}$ cool (K)      &      $3700\pm100$     & 3                     \\
L$_{cool}$  ($L\sun$)   &       $950\pm250$     & 3                     \\                  
M$_{hot}$  ($M\sun$)    &      $0.40\pm0.10$    & 3                     \\
R$_{hot}$  ($R\sun$)    &      0.04        & 3\tablenotemark{b} \\                  
T$_{eff}$ hot (103 K)   &      75            & 3                        \\
L$_{hot}$  ($L\sun$)            &      $\sim 46$     &   3\tablenotemark{c} \\
$\dot{M}$ ($M\sun$/yr)          &       $10^{-8}/6\times 10^{-8} - 10^{-7}$    & 3/4\tablenotemark{d}                   
\enddata
 \tablenotetext{a}{Calculated using Hipparcos parallax and $d=   1/\pi (")$}
  \tablenotetext{b}{ Radius value in [3] was found from R/d with d=   0.4
 kpc.  The  distance  obtained from the Hipparcos parallax (0.676 kpc) was inserted producing new radius  value.}
 \tablenotetext{c}{Calculated from $L=4 \pi R_{\rm h}^2 \sigma T_{\rm h}^4$.}
\tablenotetext{d}{Values in [4] for $d=0.4$ kpc; here re-calculated with $d=0.676$ kpc.}
\tablenotetext{e}{[1] Belczynski et al. 2000; [2] Perryman et al. 1997; 
 [3] Vogel et al. 1992; [4] {\rm Miko{\l}ajewska}, Ivison, Omont 2002}
\end{deluxetable}

We selected spectra of EG And that were closest to superior spectroscopic
conjunction, or phase 0.5, when the white dwarf is located in front of the
red giant. There were only two spectra in the IUE archive at phase 0.5.
One of them (SWP18873) was taken through the IUE small aperture, showing
the Lyman Alpha wings. However, there is one spectrum in the FUSE archive
at phase 0.5. The focus of this paper is on these three spectra.

Section 2 describes the FUV archival spectra and their preprartion for
model fitting, Section 3 describes
the modeling techniques used, and Section 4 is a discussion of our results
and conclusions.

\section{Far Ultraviolet Observations of EG And}

We used ultraviolet spectra of EG And from the IUE archive, a collection
of spectra obtained by the International Ultraviolet Explorer from 1978 to
1994.  We used Short Wavelength Prime (SWP) spectra taken at low
dispersion, with a resolution of 6\AA, through the large aperture
(10"x20"). Table 2 lists the Julian Date, date, Universal Time, phase,
exposure time in minutes, background and maximum continuum count rates,
and the observor for each spectrum.

\subsection{Characteristics of the Spectra}

The spectra of EG And exhibit a sharp rise in the FUV, especially after
being de-reddened.  We attribute this rise to the hot component of the
symbiotic system.  The spectra exhibit nebular emission lines, generally
assumed to be a result of photoionization by the accreting white dwarf.  
We studied the SWP wavelength range in order to learn more about the hot
component of each system because in the FUV, contributions from the red
giant and nebula are generally negligible.

\subsection{Flux Calibration Correction of Spectra}

Due to the amount of interstellar dust in the line of sight to the
symbiotic systems, it was necessary to de-redden the spectra before they
were analyzed.  We used E(B-V) values from Belczynski et al. (2000) as
given in Table 1.  The de-reddening process consisted of downloading the
IUE spectra in MXFITS format and correcting each data point individually
using the E(B-V) value accepted for the system.  This procedure was
carried out for each spectrum using the IDL routine UNRED.  In addition to
the reddening correction, the spectra also had to be recalibrated based
upon the Massa \& Fitzpatrick (2000) algorithms.  The Massa-Fitzpatrick
corrections are used to correct the flux calibration of the IUE NEWSIPS.
Massa and Fitzpatrick (2000) found that "the absolute flux calibration of
the NEWSIPS low-dispersion spectra was inconsistent with its reference
model and subject to time-dependent systematic effects" that could amount
to ten to fifteen percent. The flux corrections were applied before the
de-reddening process.

\clearpage 
\newpage
\setlength{\oddsidemargin}{-1.0in}
\begin{deluxetable}{lccccccc}
\tablewidth{0pc} 
\tablecaption{ EG And Observations}
\tablehead{ 
Image SWP  & 42347 &   18773 \\
Date     &  08/28/91 & 08/28/91 } 
\startdata      
JD-2440000\tablenotemark{a}&      8497.33703&      5314.60581\\
UT  &      20:05:19      &2:32:22\\
Phase  &            0.46&      0.50\\
T$_{exp}$ (min)  &      45&            15\\
Background   &20            &18\\
Max Continuum   &169            &53\\
Observor      &      M. Vogel&      R.E. Stencel\\
\enddata
\tablenotetext{a}{JDmin =   2445380, P =   482.6d (see Table 1)}
\end{deluxetable}

\clearpage 
\newpage

\setlength{\oddsidemargin}{0.0in}
\section{Synthetic Spectral Fitting}

\subsection{Construction of Photosphere and Disk Models}

To create theoretical spectra for comparison with the data, we used the
codes TLUSTY (Hubeny 1988) together with SYNSPEC (Hubeny \& Lanz 1995) to
construct model grids of atmospheric structure and absorption line
profiles for a range of temperatures and surface gravities.  Based upon
previous estimates of the stellar properties of the hot component of EG
And, we created models with T$_{eff}/1000$ from 20-150 K in increments of
10,000~K and log $g$ from 6.0 to 9.0 in increments of 0.5.  All models
assumed solar abundances and a rotational velocity of 100 km/s.  Each of
the spectra was also compared to models computed with TLUSDISK. These
accretion disk models had a range of white dwarf masses, accretion rates,
and inclination angles (See Wade \& Hubeny 1998 for a detailed description
of the modeling procedure and a table of models).  We compared our spectra
to all of the models discussed in their paper with the exception of those
models with an inclination of $i=8^{\circ}$ because they consistently
produced spectra that bore no semblance to a normal spectrum.

The spectra we examined had strong emission lines, which are generally
assumed to result from photoionization of the red giant wind material by
the accreting hot component.  It was necessary to mask out these emission
regions in the observations in order to get a more accurate fit to the
continuum.  Table 3 lists the emission line regions that were masked for
the spectra.

\begin{deluxetable}{ccc}
\tablewidth{0pc} 
\tablecaption{Table 3. Emission Regions Masked: EG And (all)}
\tablehead{
\colhead{Spectra}&
\colhead{Regions Masked (\AA)}&\nodata}
\startdata
1170-1185,& 1197-1227,& 1231-1250,\\
1290-1320, &1384-1413,& 1475-1485,\\
1534-1563,& 1628-1654,& 1715-1755,\\
1880-1990&&\\
\enddata
\end{deluxetable}

We used our fitting routine, IUEFIT, to generate an error bar that was
10\% of the flux.  We took the error dispersion to be this error bar plus
a small portion of flux to avoid giving too much weight to the low flux
levels. For each trial photosphere or accretion disk model, a reduced
$\chi^{2}$ value and scale factor were generated. The resulting distance
was then compared to the parallax-derived distance for EG And.

\subsection{LTE Photosphere Fits}

We constructed grids of models for both the LTE and NLTE cases. In all
cases, the same masking of emission lines was used in the preparation of
the observed spectrum.  The best-fitting LTE photosphere fit of SWP42347
had a temperature of 50,000K, log g = 6.5, lower than the temperature of
75-90,000K suggested in the literature (see Table 1) and determined from
modified Zanstra techniques.

\subsection{NLTE Photosphere Fits}

We constructed a grid of NLTE photosphere models by using the NLTE option
in TLUSTY198. We selected the spectra SWP42347 and SWP18773 for this
comparison using the same masking of emission line regions as in the LTE
fit.  For the NLTE grid, we computed models for the following parameter
ranges: T$_{eff}$/1000 from 40 to 90K in steps of 1000K, log $g$ from 6.0
to 8.5 in steps of 0.5, and assumed solar composition for all models.
Based upon the lowest $\chi^{2}$ value (1.8214), we found the best-fitting
NLTE model to SWP42347 to have T$_{eff} = 41,000$K with log $g = 6.5$ with
a scale factor $S= 1.371\times 10^{-2}$. This NLTE fit is displayed in
figure 1. For log $g= 6.5$ (a high gravity subdwarf) we take
$R_{wd}/R_{\sun} = 2.50\times 10^{-2}$, which implies a distance d = 250
pc. For the small aperture spectrum SWP18773, which appears to show the
Lyman $\alpha$ wings with minimal geocoronal emission contamination, we
obtain T$_{eff} =50,000$K with log $g = 8.5$. While the best-fit gravity
corresponds to a more massive white dwarf, it is remarkable that this
model fit gives a scale factor distance much closer to the parallax value.
This NLTE fit is displayed in Figure 2.

An archival FUSE spectrum of EG And taken at phase 0.46 allows an
additional, independent determination of temperature. We carried out NLTE
model photosphere and accretion disk analyses of this FUSE spectrum,
taking the same parameter range that we explored in the IUE archival
analysis. Our modeling of the FUSE data yielded, for a single temperature
white dwarf, a best-fitting temperature T$_{eff} = 50,000$K in excellent
agreement with the IUE temperature.

Nevertheless, since the hot component is widely believed to have a low
mass (Vogel et al. 1992; Muerset et al. 2000), and the orbital inclination
is known to be high, this offers the advantage of reducing the number of
free parameters in our fitting by fixing the mass of the white dwarf at
0.4 Msun with a radius of 0.04 Rsun and the orbital inclination at 75
degrees, as given in table 1 from the refereed literature. For this mass,
log g = 6.84. We also fixed the distance at 676 pc, the distance given by
the Hipparcos parallax, and took the abundances to be solar and the
rotational velocity to be 100 km/s.  The masking and the binning of the
FUSE spectrum was the same as above. Likewise, for the accretion disk
model fits, we took M$_{wd} = 0.4 M_{\sun}$. and fixed the inclination at
the published value of 75 degrees and fixed the distance at 676 pc.

The best-fitting NLTE white dwarf model yielded T$_{eff} = 50,000$K with a
reduced $\chi^{2}$ value = 4.032 and is displayed in figure 3. Once again,
this value of T$_{eff}$ agrees with the temperature we derived from the
IUE spectra at superior spectroscopic conjunction when the white dwarf is
in front of the red giant.

\subsection{ Accretion Disk Fits}

The accretion disk models provide an interesting initial exploration of
the source of FUV flux from the EG And hot component. Therefore, we
carried out fits to the three spectra at 0.5 phase, the IUE spectra
SWP18773 and SWP42637 and the FUSE spectrum. Once again, we fixed the
white dwarf mass at $M_{wd} = 0.4 M_{\sun}$, the inclination angle
$75^{\circ}$ and adopted the distance of 676 pc given by the Hipparcos
parallax. By comparison, using the parameters from Table 1, we find the
best-fitting optically thick accretion disk model to the FUSE spectrum
have $\dot{M} = 1\times 10^{-8}$ M$_{\sun}$/yr but the $\chi^{2}$ value =
31.45. As shown in figure 4, at an inclination of 75$^{\circ}$, the disk
flux at $10^{-8} M_{\sun}$/yr is far too low to match the observed FUSE
flux level. At least an order of magnitude higher accretion rate would be
required to reach the observed FUSE flux level. An accretion rate as high
as $10^{-7} M_{\sun}$/yr is implausible in view of the observed red giant
mass loss rate (see Table 1) and the predicted accretion efficiency from
3D hydrodynamic simulations of wind accretion using symbiotic parameters.
The best-fitting accretion disk models to the IUE spectra SWP42347 and
SWP18773 and the FUSE spectrum are summarized in Table 3.
    
\begin{deluxetable}{cccccc}          
\tablewidth{0pc} 
\tablecaption{Best Disk Fit Parameters}
\tablehead{ 
\colhead{SWP }&\colhead{Mass($M\sun$)}&\colhead{$Log \dot{M}({M}\sun/yr)$}&\colhead{Inclination$^{\circ}$}&\colhead{ Distance (pc)}&\colhead{$\chi^{2}$} }
\startdata
42347 &      0.4&       -8.0 &       75$^{\circ}$&   160          &
0.239764\\
18773 &     0.4&       -8.0&         75$^{\circ}$&  181     & 2.02664
\\
FUSE    &    0.4&       -8.0&         75$^{\circ}$&            & 31.45  \\
\enddata
\end{deluxetable}

\section{Conclusions}

We have found that for the widely accepted parameters of EG And, i.e., a
low hot component mass of $0.4 M_{\sun}$, a system inclination of 75
degrees and a distance of 676 pc parallax, the hot component has a
temperature of 50,000K and is the principal source of the FUV continuum in
the IUE spectra and FUSE spectrum taken at orbital phase 0.5. On the other
hand, accretion disk models at moderately high inclinations with accretion
rates $\dot{M} = 1\times 10^{-8}$ for white dwarf masses $M_{wd} = 0.4
M_{\sun}$ are grossly inconsistent with the distance from the Hipparcos
parallax. This is surprising since, on theoretical grounds, permanent,
stable disks are predicted to form from red giant wind mass transfer to
the white dwarf (Mastrodemos and Morris 1998). It is important to note
however that in the Mastrodemos and Morris (1998) simulations, the effects
of the radiation field of the hot component and any hot component magnetic
field are neglected. Moreover, our finding is consistent with other
studies which rule out a disk independently based upon spectroscopic
evidence. Our study is the first to do so using actual optically thick
disk models with vertical structure.

The phases near phase 0.5 are far outside the range where Rayleigh
scattering would have an appreciable effect and attenuate the continuum.
We regard the lower photospheric temperatures as potential indications
that the real surface temperatures may be substantially lower than the
Zanstra values. A similar conclusion was reached for the S-type system RW
Hydrae by Sion et al. (2002).

\acknowledgments

This research was supported in part by NSF grant 99-01195, NASA ADP grant
NAG5-8388 (EMS), by summer research funding from the NASA-Delaware Space
Grant Colleges Consortium (KK, JM) and by the Polish KBN grant No. 5P03D
019 20 (JM).

Figure Captions

Fig.1-- The flux (ergs/cm$^{2}$/s/\AA) versus wavelength (\AA) for the IUE    
spectrum SWP42347 of EG And along with the best-fitting NLTE model to
the observed continuum with all strong emission lines masked in the fit.    
The resulting temperature is 41,000K, with log $g =   6.5$
and solar composition. The best-fitting LTE model to this same spectrum
had a temperature of 50,000K.

Fig.2-- The flux (ergs/cm$^{2}$/s/\AA) versus wavelength (\AA) for the IUE
spectrum SWP18773SM of EG And taken through the small IUE aperture along
with the best-fitting NLTE model to the observed continuum with all strong
emission lines masked in the fit. The resulting temperature is 41,000K,

Fig.3-- The flux (ergs/cm$^{2}$/s/\AA) versus wavelength (\AA) for an
archival FUSE spectrum taken near orbital phase 0.5, along with the
best-fitting NLTE photosphere model to the observed continuum with all
strong emission lines masked in the fit. The hot component mass was fixed
at the published value of $M_{wd} = 0.4 M_{\sun}$ and the distance fixed
at 676 pc. The resulting best-fitting temperature is 50,000K.

Fig.4-- The flux (ergs/cm$^{2}$/s/\AA) versus wavelength (\AA) for the
same archival FUSE spectrum taken near orbital phase 0.5, along with the
best-fitting accretion disk model to the observed continuum with all
strong emission lines masked in the fit. The hot component mass was fixed
at the published value of $M_{wd} = 0.4 M_{\sun}$, the orbital inclination
fixed at the published value of 75$^{\circ}$. In order for an optically
thick accretion disk to fit the FUSE spectrum, an accretion rate greater
than $10^{-7} M_{\sun}$/yr is required which is implausible in view of the
known red giant mass loss rate and the accretion capture efficiency.


\begin{thebibliography}{}
\bibitem[]{}
Belczynski, K., Mikolajewska, J., Munari, U., Ivison, R.J., \& Friedjung,
    M.2000, A\&AS, 146, 407

\bibitem[]{}
Hubeny, I. 1988, Comput. Phys. Commun., 52, 103

\bibitem[]{}
Hubeny, I. \& Lanz, T. 1995, ApJ, 439, 875

\bibitem[]{}
Hubeny, I. \& Lanz. T. 1995 GSFC Tlusdisk- A User's Guide
\bibitem[]{}

\bibitem[]{}
Massa, F. \& Fitzpatrick, E.L. 2000, ApJS, 126, 517

\bibitem[]{}
Mastrodemos, N. \& Morris, M.1998, ApJ, 497, 303

\bibitem[]{}
Miko{\l}ajewska et al. 2002, Adv. Space Res. 30, 2045

\bibitem[]{}
Muerset, U., Nussbaumer, H., Schmid, H.M., \& Vogel, M. 1991, A\&A, 248,    
458

\bibitem[]{}
Muerset, U., Dumm, T., Isenegger, S., Nussbaumer, H., Schild, H., Schmid,
H.M., \& Schmutz, W.2000, A\&A, 353, 952

\bibitem[]{}
Perryman, M.A.C., Lindegren, L., Kovalevsky, J., Hog, E., Bastian, U.,    
Bernacca,    
  P.L., Creze, M., Donati, F., Grenon, M., Grewing, M., Van Leeuwen, F.,    
Van Der Marel, H.,    
  Mignard, F., Murray, C.A., Le Poole, R.S., Schrijver, H., Turon, C.,    
Arenou, F.,    
  Froeschle, M., \& Petersen, C.S. 1997, A\&A, 323, 49

\bibitem[]{}
Schild, H. et al. 1992; MNRAS 258, 95

\bibitem[]{}
Sion, E.M., Mikolajewska, J., Bambeck, D., \& Dumm, T. 2002, AJ, 123, 983

\bibitem[]{}
Vogel, M., Nussbaumer, H., \& Monier, R. 1992, A\&A, 260, 156

\bibitem[]{}
Wade, R.A. \& Hubeny, I. 1998, ApJ, 509, 350

\bibitem[]{}
Wilson, R.E. \& Vaccaro, T.R. 1997, MNRAS, 291, 54

\end{thebibliography}
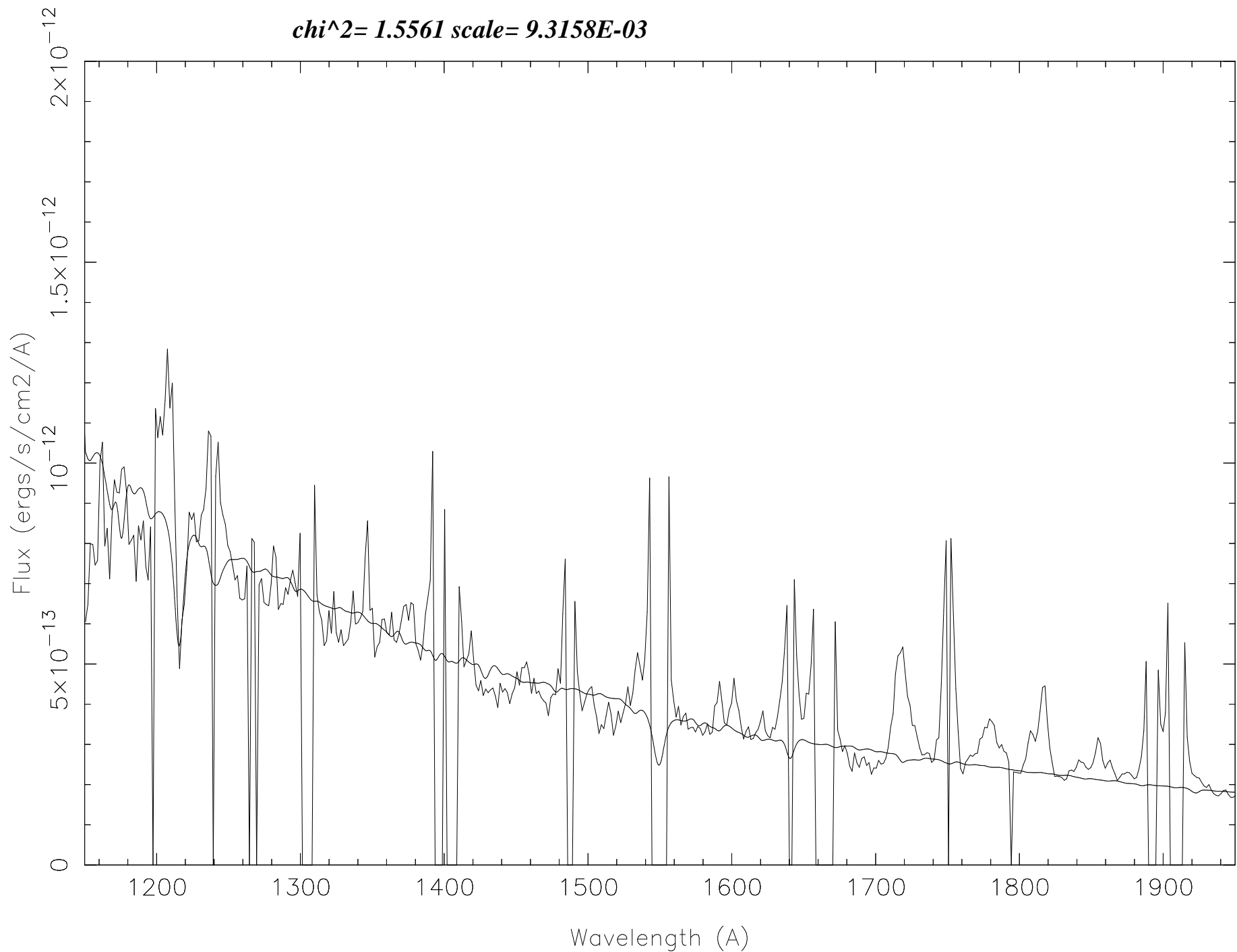
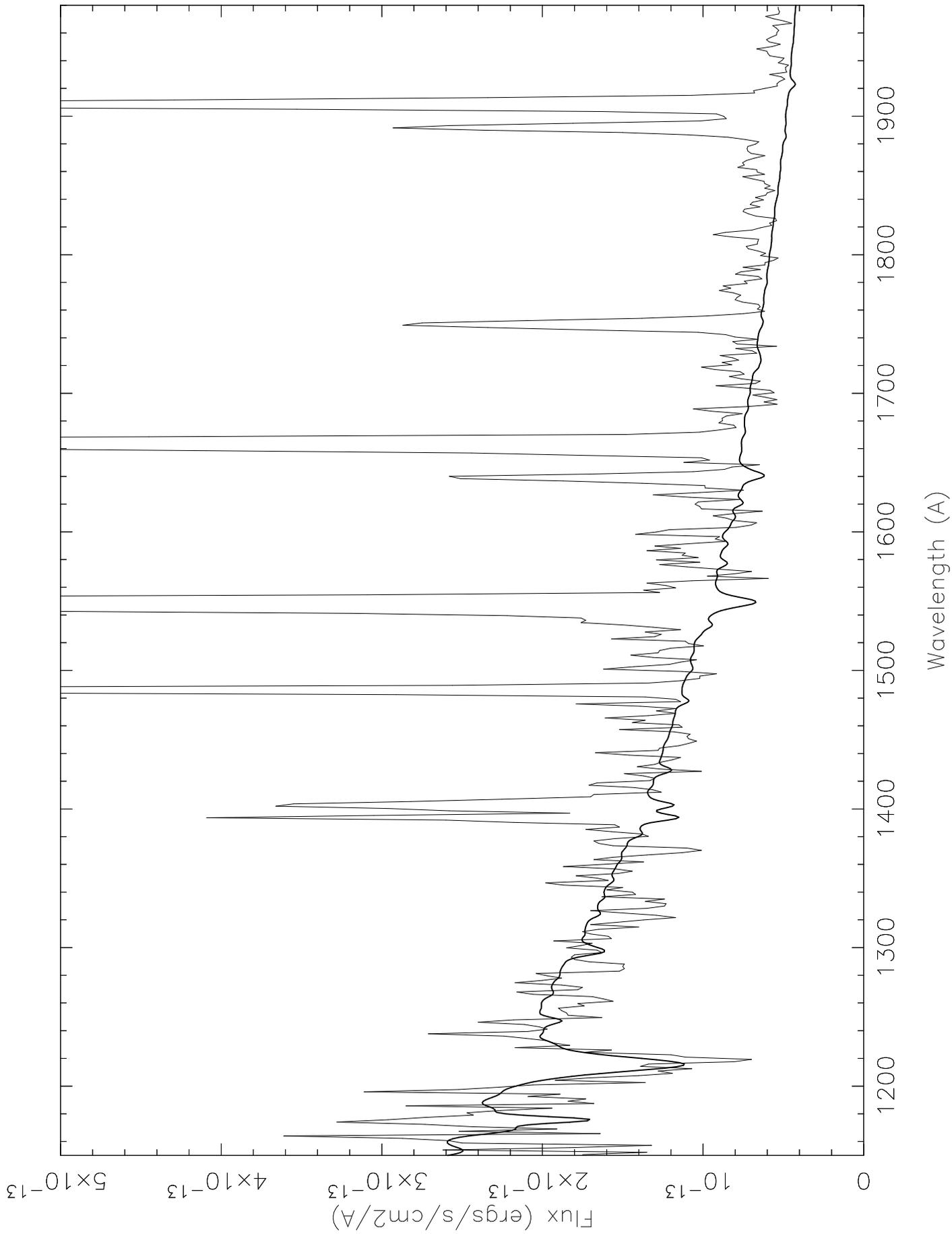
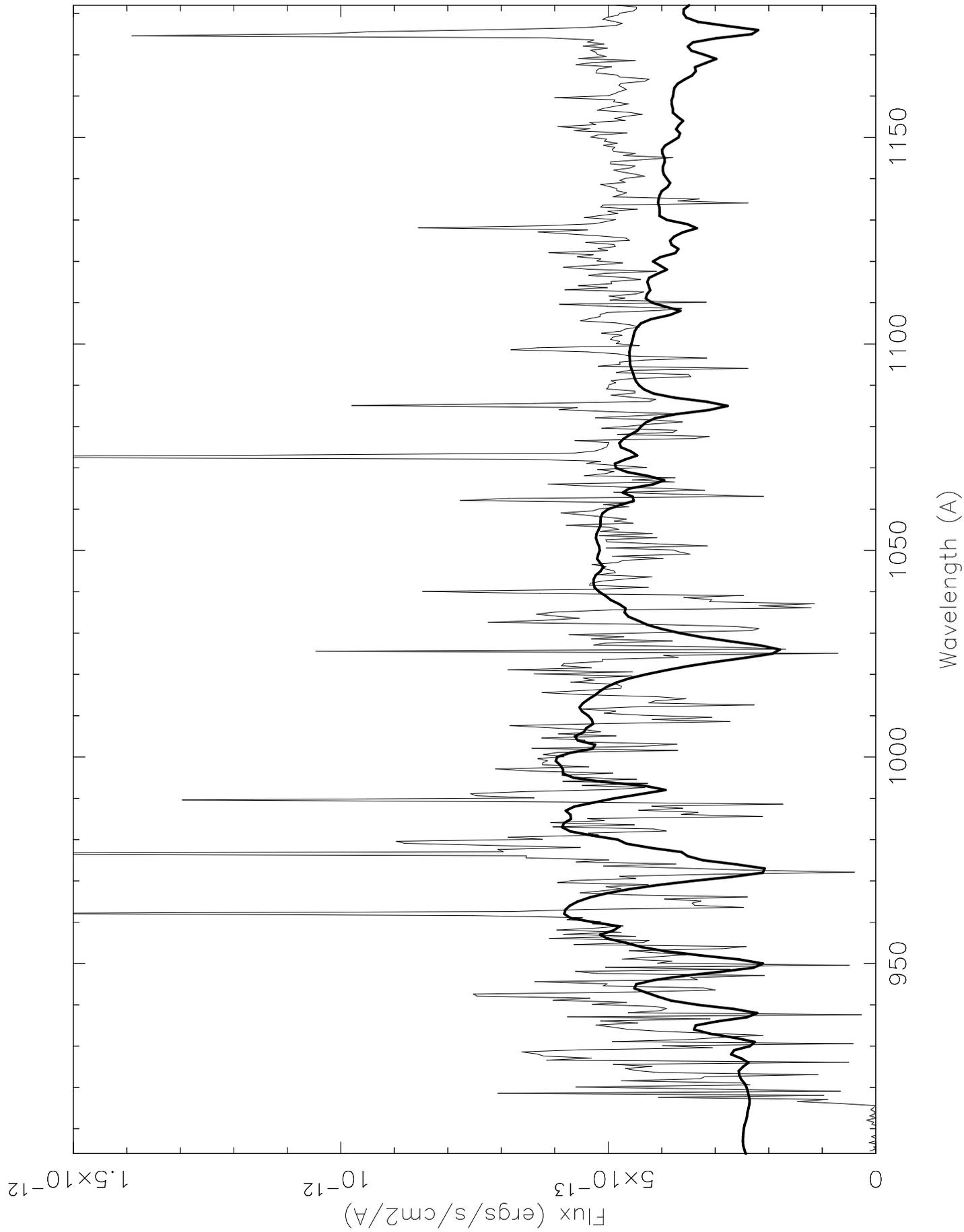
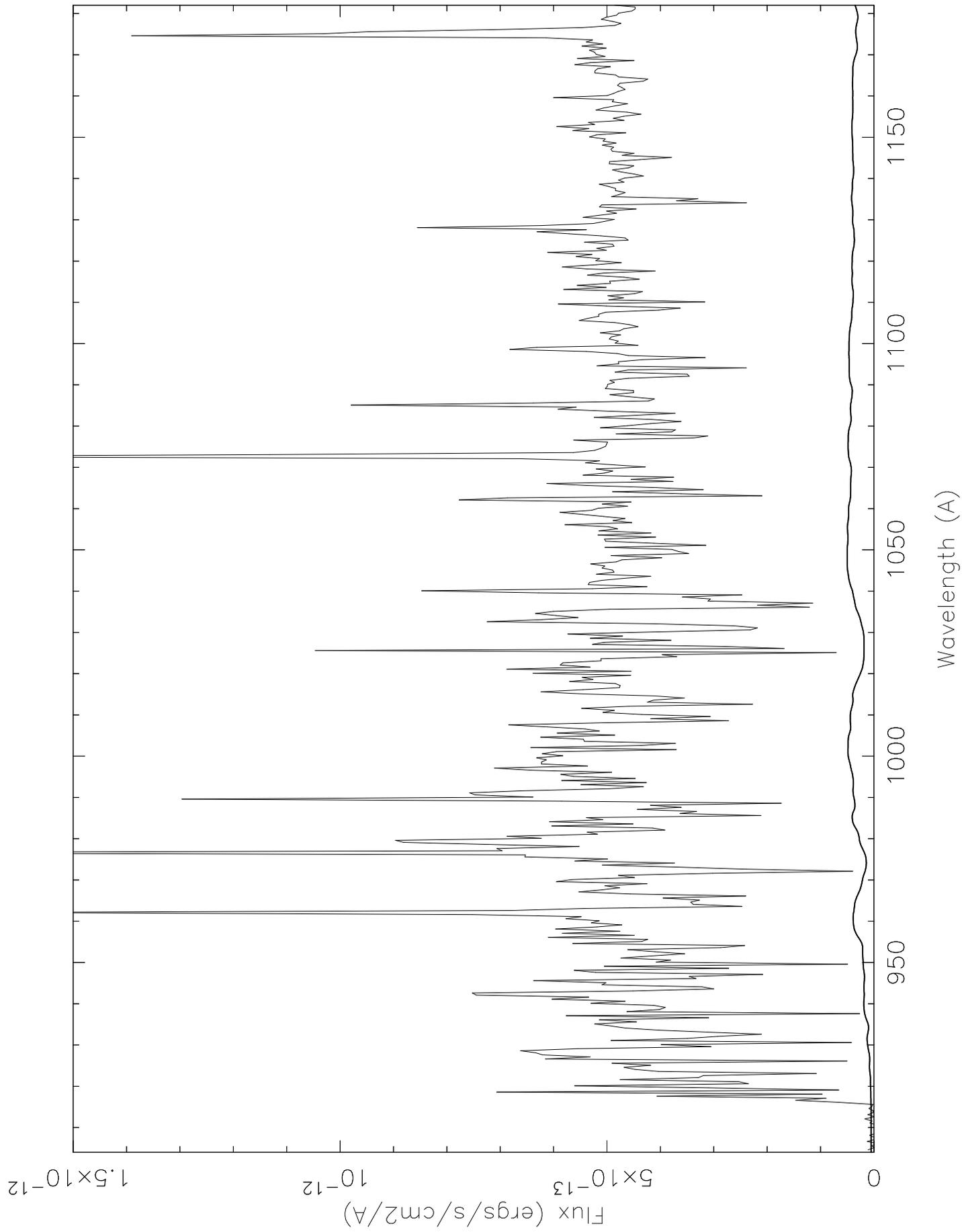
\end{document}